\documentclass[doublecol]{epl2} 
\usepackage{graphicx}
\usepackage{amssymb}
\usepackage{amsmath}
\usepackage{amsfonts}

\title{Micro-flock patterns and macro-clusters in chiral active Brownian disks}

\shorttitle{Liquids 2017}

\author{Demian Levis \inst{1}  \inst{2}  \inst{3} \and Benno Liebchen \inst{4}  \inst{5}}

\institute{
\inst{1} Department de F\'isica de la Mat\`eria Condensada, Universitat de Barcelona, 
Mart\'i i Franqu\`es 1, E08028 Barcelona, Spain \\
\inst{2}  UBICS University of Barcelona Institute of Complex Systems, 
Mart\'i i Franqu\`es 1, E08028 Barcelona, Spain \\
\inst{3} CECAM Centre Europ\'een de Calcul Atomique et Mol\'eculaire, \'Ecole Polytechnique F\'ed\'erale de Lausanne, Batochimie, Avenue Forel 2, 1015 Lausanne, Switzerland \\
\inst{4} SUPA, School of Physics and Astronomy, University of Edinburgh, Peter Guthrie Tait Road, Edinburgh, EH9 3FD, UK \\
\inst{5} Institut f\"{u}r Theoretische Physik II: Weiche Materie, Heinrich-Heine-Universit\"{a}t D\"{u}sseldorf, D-40225 D\"{u}sseldorf, Germany
}

\abstract{
Chiral active particles (or self-propelled circle swimmers) feature a rich collective behavior, comprising rotating macro-clusters and micro-flock patterns which consist of phase-synchronized rotating clusters 
with a characteristic self-limited size. 
These patterns emerge from the competition of alignment interactions and rotations suggesting that they might occur generically in many chiral active matter systems. 
However, although excluded volume interactions occur naturally among typical circle swimmers, 
it is not yet clear if macro-clusters and micro-flock patterns survive their presence. 
The present work shows that both types of pattern do survive but feature strongly enhance fluctuations regarding the size and shape of the individual clusters.
Despite these fluctuations, we find that the average micro-flock size still follows the same characteristic scaling law as in the absence of excluded volume interactions,
i.e. micro-flock sizes 
scale linearly with the single-swimmer radius.}
\begin{document}

\maketitle

\section{Introduction}\label{sec:intro}
Understanding the statistical mechanics of circle swimmers represents an important but relatively new
problem in active matter physics. This new active matter class, now going under the name of 'chiral active matter', involves many biological micro-swimmers ranging 
from bacteria like Escherichia Coli \cite{Berg1990, diLuzio2005, Lauga2006, DiLeonardo2011}
and Listeria monocytogenes \cite{Shenoy2007} to spermatozoa \cite{Friederich2012, Riedel2005}, all of which swim circularly (or helically) rather than along straight trajectories. 
In fact, the ubiquity of chiral active swimmers in the microbiological world is not a surprise, but rather the consequence of a general principle which can be illustrated as follows: 
conversely to a passive rigid body in a constant force field, active particles produce a force (in the 'dry' picture) which pushes them 
forward and points in a constant direction in the body frame; in the lab frame, the forcing direction evolves along with the particle.
Generically, an active particle swims linearly only if it features a common symmetry axis of body (shape) and self-propelling force. Otherwise it experiences a 
constant torque leading to circular motion.
Linear swimmers can thus be seen as the exception rather than the rule \cite{Loewen2016}. 
Following this symmetry-principle it has been possible to also design synthetic circle swimmers of e.g. L-shaped bodies \cite{Kummel2013, Hagen2014}.

Chiral active matter not only encompasses a large class of biological and synthetic micro-swimmers, but 
also creates a plethora of new phenomena beyond the physics of linear active matter. Examples include
vortex like (macro)clusters of curved polymers \cite{Denk2016}, rotating doublets \cite{Kaiser2013}, spiral waves \cite{Liebchen2016} and whole patterns consisting of rotating micro-flocks \cite{CAP}.
These micro-flocks 
are phase-synchronized clusters of rotating particles emerging at a characteristic self-determined size, which emerge as a consequence of flocking (as in the Vicsek model \cite{Vicsek1995}) in a chiral system.
Although micro-flocks are often much larger than the single swimmer radius, their size scales linearly with it, which should allow to control the pattern, e.g. via the properties of a single swimmer.  

While micro-flock patterns emerge as a direct consequence of the interplay between self-propulsion and alignment interaction among circle swimmers \cite{CAP} - neither require attractive nor repulsive interactions 
to emerge - it is important to understand their robustness against excluded volume interactions which are naturally present in any conceivable circle swimmer ensemble.
Such excluded volume interactions sometimes act as a key player in active matter: for motility-induced phase separation (MIPS) in particular, 
they block oppositely moving active particle such that they stay together before rotational diffusion releases them \cite{Buttinoni2013}; this mechanism can eventually trigger a  
phase separation for large enough density and self-propulsion velocity  \cite{Redner2013, Stenhammar2014, Cates2015, Levis2017}. 

In the present work, we explore the impact of excluded volume interactions on the collective behavior of chiral active matter. 
Here we are specifically interested in the robustness of micro-flock pattern formation -- and look for new excluded-volume-induced effects on the pattern morphology. 
As a key result, we find that micro-flock patterns do not only survive the presence of excluded volume interactions, but find that even their characteristic scaling with the single circle swimmer radius persists.
The morphology of the patterns, however, experiences enhanced fluctuations regarding the cluster-shape and size in presence of the interactions, such that the overall pattern is more heterogeneous than in the noninteracting case.  
We also find that the phase-synchronization of particles within each micro-flock persists in principle 
but features a broader distribution of phases around the average than in absence of excluded volume interactions. 

\section{Model and methods}\label{sec:model}

We model a suspension of circle-swimmers as a set of $N$ identical self-propelled particles that swim with a constant speed $v_0$ along a direction $\boldsymbol{n}_i=(\cos\theta_i,\sin\theta_i)$ in the Euclidean plane. 
This direction evolves by diffusion, by the influence of a constant torque acting on the particles (which is responsible for circular swimming) and by the mutual interaction between neighboring swimmers. 
The governing equations of motion are
\begin{equation}
\dot{\boldsymbol{r}}_i(t)= v_0 \boldsymbol{n}_i(t) + \mu \boldsymbol{F}_i
\end{equation}
\begin{equation}\label{eq:angle}
\dot{{\theta}}_i(t) = \omega +\frac{K}{\pi R^2}\sum_{j\in\partial_i}\sin(\theta_j-\theta_i)+\sqrt{2D_r}{\eta}_i(t)
\end{equation}
Here $\boldsymbol{r}_i(t)$ denotes the spatial location of the $i$-th circle-swimmer at time $t$ and $\mu$ is its mobility. 
The forces $\boldsymbol{F}_i$ account for excluded volume interactions between swimmers and are defined as $\boldsymbol{F}_i=-\nabla_i\sum_{j<i} U(|\boldsymbol{r}_i-\boldsymbol{r}_j|)$ where
\begin{equation}
 U(r)=u_0\left(\frac{\sigma}{r}\right)^{12}\, 
\end{equation}
with an upper cutoff at $3\sigma$. The parameters $\sigma$ and $u_0$ provide the natural length and energy scale of the model.  
Besides the excluded volume effects described by the short-range repulsive potential, swimmers are also subject to velocity alignment interactions, akin to the ones defining the Vicsek model \cite{VicsekRev}. 
Alignment interactions are introduced  as a torque in our over-damped formulation eq. (\ref{eq:angle}), and their strength is controlled by the coupling constant $K\geq0$. 
The sum eq. (\ref{eq:angle}) runs over nearest neighbors defined by the interaction range $R_{\theta}$ (only particles distant of less than $R_{\theta}$ align each other). 
The term led by the coefficient $D_r$ describes rotational diffusion; here $\eta_i$, represents zero-mean and unit variance Gaussian white noise. 
In the absence of repulsive interactions (i.e. $u_0\to 0$) we recover the chiral active particle model discussed in \cite{CAP}, which we extend here to 
consider disks of finite size experiencing crowding effects.   

After fixing $R_{\theta}=2\sigma$, we identify the following set of dimensionless parameters: the particle density $\rho=N\sigma^2/L^2$, 
the normalized rotational frequency $\Omega=\omega/D_r$, the reduced coupling parameter $g=\frac{K}{4 \pi \sigma^2 D_r}$ 
and the P\'eclet number Pe$=\frac{v_0}{\sigma D_r}$. From now on, we express lengths in units of $\sigma$ and time in units of $D_r^{-1}$ and 
fix $\rho=0.1$, $u_0=1$ so that Pe, $\Omega$ and $g$ serve as our dimensionless control parameters. 
Here, the choice of $\rho=0.1$ and Pe$\leq 20$ ensures that the system is well below the regime where structure formation via the MIPS-mechanism 
might occur even in the absence of 
circular swimming. Therefore, all the structures we describe below emerge due to the competition between alignment interactions and rotations only, 
but once emerged, they may be affected by excluded volume interactions, as we shall see below in more detail. 

\section{Flocking and pattern formation}\label{sec:flock}

In the absence of rotations ($\Omega=0$) and excluded volume interactions ($u_0=0$), the competition between velocity alignment and noise gives rise to the so-called flocking transition, 
extensively discussed in the context of Vicsek-like models \cite{Toner1995, Toner2005, VicsekRev}. Above a critical coupling $g^*$, global polar order 
emerges in the system: the particles move collectively, on average, along a given direction. 
Remarkably, this symmetry breaking phase transition is accompanied by structure formation in the form of traveling bands \cite{Chate2008}. 

Presence of rotations ($\Omega>0$) strongly alters this behavior \cite{CAP}. 
While rotations do not affect the location of the flocking transition of point-like polar particles, they qualitatively change the ordered phase \cite{CAP}: 
slow rotations  ($\Omega \lesssim 1$) induce large rotating aggregates with droplet like shapes which coarsen and ultimately phase separate.
Faster rotations  ($\Omega \gtrsim 1$) frustrate the tendency of the particles to align their swimming direction, giving rise to a pattern of micro-flocks, i.e. phase-synchronized rotating aggregates, with a characteristic 
self-limiting size \cite{CAP}. 
Since the onset of flocking $\rho g^*=2$ in the standard Vicsek model is essentially invariant against both, rotations \cite{CAP} and density-dependent ('quorum sensing'-like) interactions
\cite{Farrell2012}, we expect that the transition to the ordered phase in the CAP model still occurs at $\rho g^*\approx 2$ if rotations and excluded volume interactions are present simultaneously.

To specify these expectations and to study the impact of excluded volume interactions on macrocluster formation and microflock patterning in detail, 
we now perform Brownian dynamics simulations of $N=2000$ up to $N=16000$ circularly swimming disks in a quadratic box with periodic boundary conditions. 

\begin{figure}[h]
\begin{center}
\includegraphics[scale=0.48,angle=0]{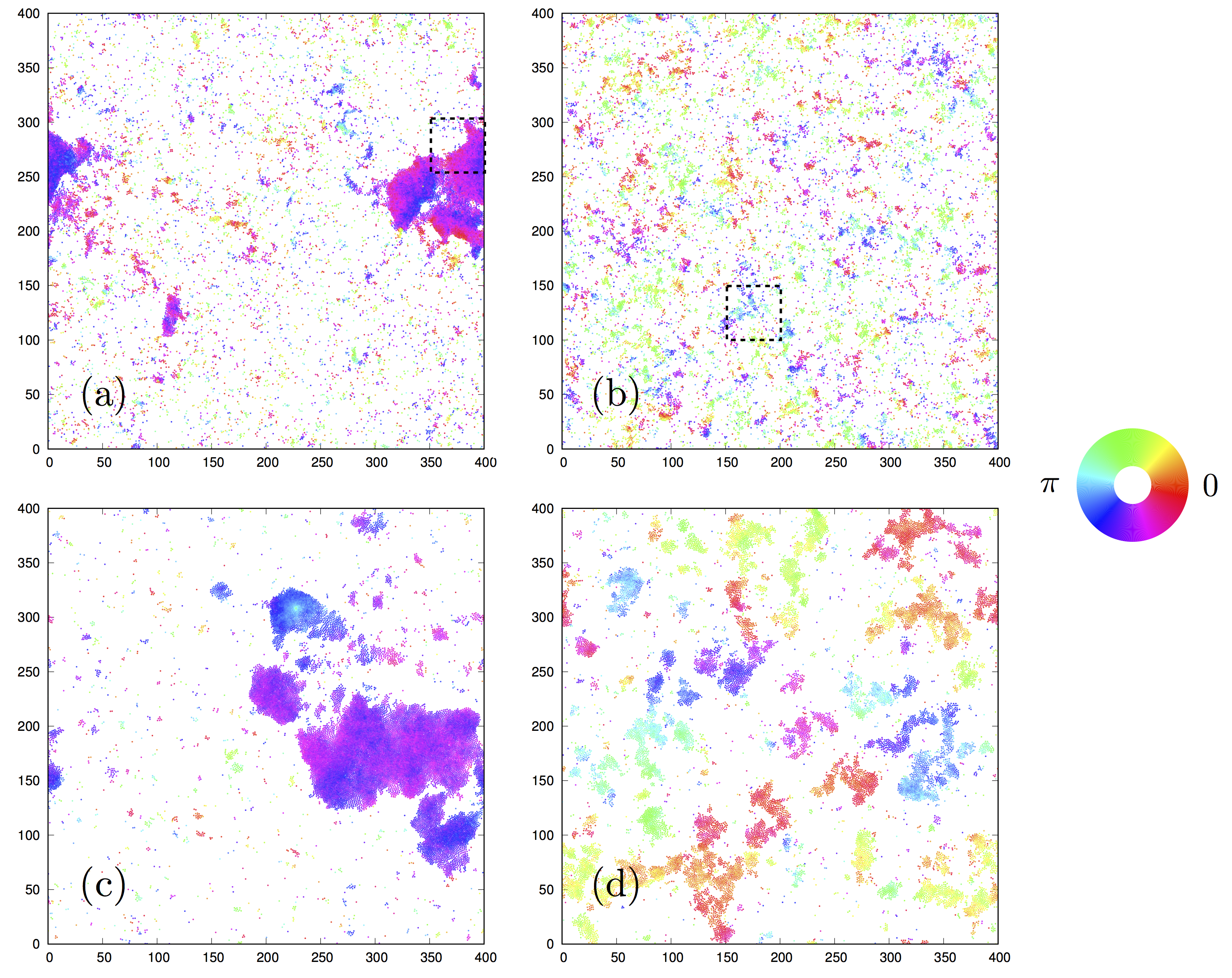}
\caption{Steady-state configurations for systems made of $N=16000$ circle-swimmers at a density $\rho=0.10$ and Pe$_r=10$. 
A color map is associated to the swimmers' orientation $\{\theta_i\}$. (a): $g\rho=3$, $\Omega=0.2$; (b):  $g\rho=3$, $\Omega=3$; (c): $g\rho=20$, $\Omega=0.2$; (c): $g\rho=20$, $\Omega=3$. }
\label{fig:snaps}
\end{center}
\end{figure}

In Fig. \ref{fig:snaps} we show representative snapshots for $N=16000$ chiral active disks. 
For slow rotations ($\Omega=0.2$), we observe a 
large high density polar structure, or \emph{macro-drop}, coexisting with a disordered background at lower density. As for $u_0=0$, this behavior 
resembles the phase separation accompanying a liquid-gas phase transition. As the coupling strength increases (see Fig. \ref{fig:snaps} (a), (c) for $g\rho=3$, $20$) the 
density difference between the macro-drop and the surrounding gas is accentuated. In any case, this structure rotates coherently and features large fluctuations. 
The drop breaks and re-shifts in short time scales, showing that its structure is subject to much larger fluctuations that its counterpart in the absence of excluded volume (see Supplementary Movie 1 \cite{SM}). 

For faster rotations ($\Omega=3$), a new pattern consisting of comparatively small rotating clusters, similar to the micro-flocks reported in  \cite{CAP}, appears (see Fig.~\ref{fig:snaps} and Supplementary Movie 2\cite{SM}). In this high frequency regime, clusters 
do not coalesce or coarsen within the explored (long) time-scales, but show a 
self-limited size, which can be controlled by the parameters determining the locomotion of the individual particles (see discussion below). 
While each cluster, or \emph{micro-flock}, displays polar order (encoded with colors in Fig.~\ref{fig:snaps}), the overall system does not exhibit global order. 
Compared to the patterns explored for point-like circle swimmers \cite{CAP} the shape of the present structures is far more irregular (see Supplementary Movie 2\cite{SM}).
While, the mechanism leading to these patterns requires alignment interactions and rotations only, for repulsive disks, multi-particle collisions introduce additional 
fluctuations and packing constraints which are responsible for the comparatively irregular structures. 
It is interesting to compare the present (fluctuating) structures with the somewhat reminiscent patterns observed in 2D suspensions of sperm cells \cite{Riedel2005} and protein filaments \cite{Loose2013}.
In the following sections we describe and discuss in more detail the nature of the macro-drops and the 'fluctuating micro-flocks'.

\section{Macro-drop regime}\label{sec:phasesep}
The characteristic feature of the macro-drop regime is the emergence of a dense cluster whose size scales linearly with the system size (at late times). 
Here, we define a cluster as a set of connected disks with inter-particle distance smaller that $R_{\theta}$, the polar interaction radius. We then measure the 
cluster mass distribution $P_m$ as the normalized distribution measuring the frequency of occurrence of clusters made of $m$ disks. Several quantities can be extracted from the cluster analysis. 
In particular, we can identify the largest cluster $c$ in the system and measure its radius of gyration as 
\begin{equation}
R_g=\sqrt{\frac{1}{m_{c}}\sum_{i\in c}|\boldsymbol{r}_i-\boldsymbol{r}_c |^2}\, , 
\end{equation}
where the sum runs over all the disks belonging to the largest cluster, of size $m_c$ and with center of mass $\boldsymbol{r}_c$. The radius of gyration provides a characteristic length 
scale associated to the macro-drop. Note that other length scales could have been defined, using the mean cluster size, or density correlations. The drawback of these latter quantities is that 
the low-density phase is also taken into account and are therefore more noisy that $R_g$, which directly focuses on the relevant structure. 

\begin{figure}
\begin{center}
\includegraphics[scale=0.45,angle=0]{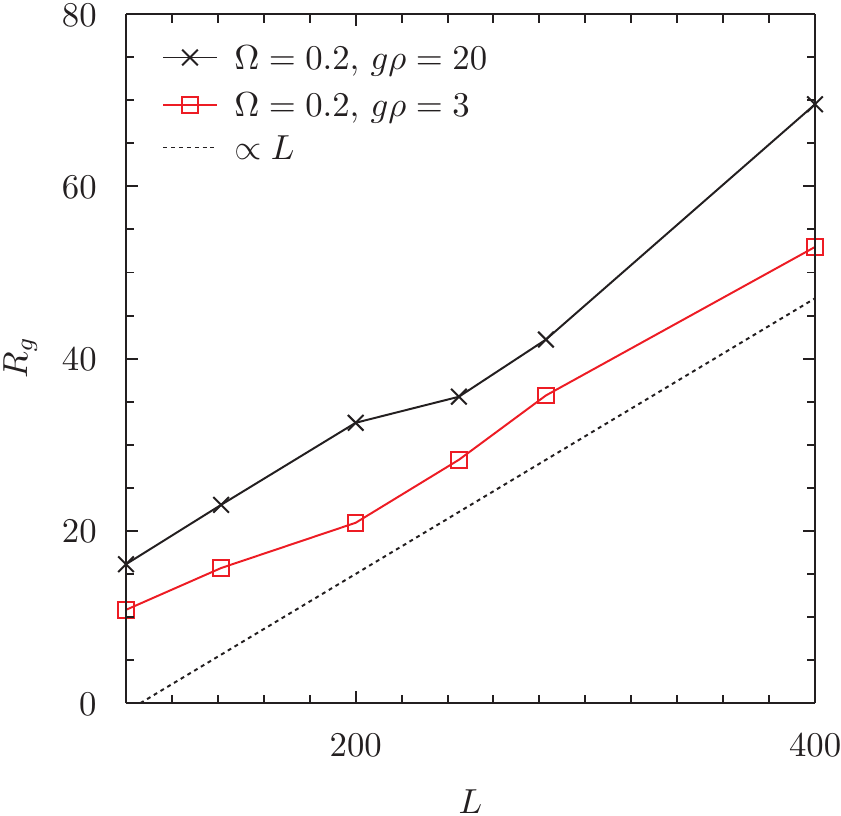} \includegraphics[scale=0.45,angle=0]{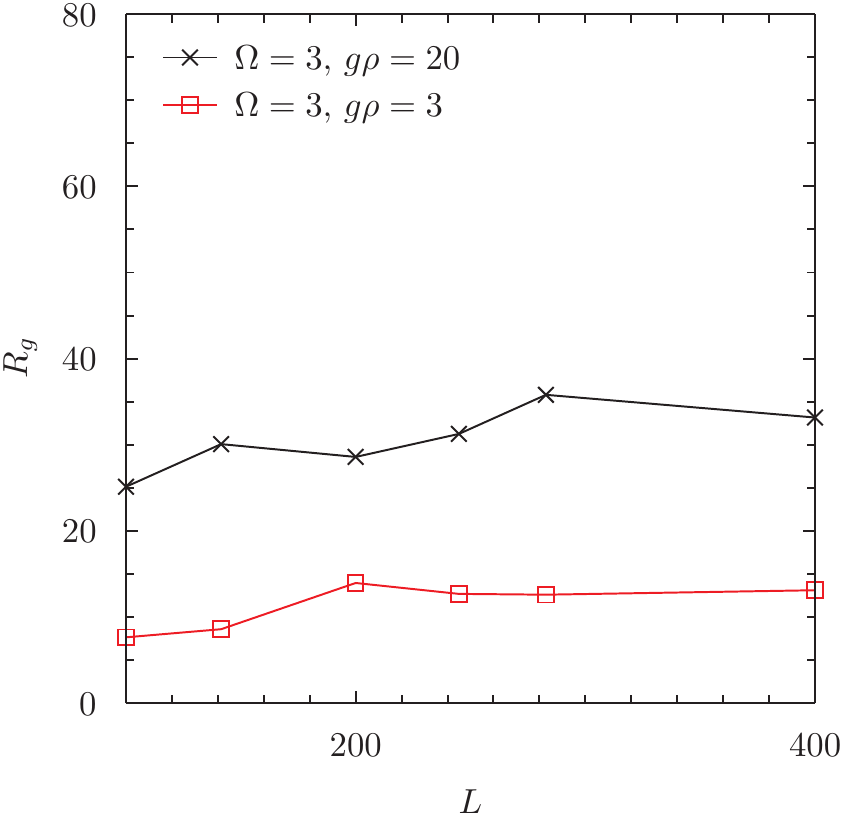}
\caption{Characteristic length-scale $R_g$ as a function of system size in the macro-drop ($\Omega =0.2$) (left) and micro-flock ($\Omega =3$) (right) regime for Pe$=10$. }
\label{fig:Rg}
\end{center}
\end{figure}

As shown in Fig. \ref{fig:Rg}, the size of the largest cluster, i.e. the drop, grows proportionally with the system size $L$ suggesting the presence of phase separation.
In fact, at a given value of the parameters, the system selects the density of the coexisting phases. As the interaction parameter (or density) increases, the density of the dilute region 
decreases and the size of the drop grows. 
This is different to what happens in the micro-flock regime: there, the size of the structures emerging for faster rotations does not scale with the system size but is self-limited. 
As shown in Fig. \ref{fig:Rg}, $R_g$ is roughly independent of $L$, showing that clusters in this regime have a characteristic size and thus constitute a proper pattern. 

\begin{figure}
\begin{center}
\includegraphics[scale=0.45,angle=0]{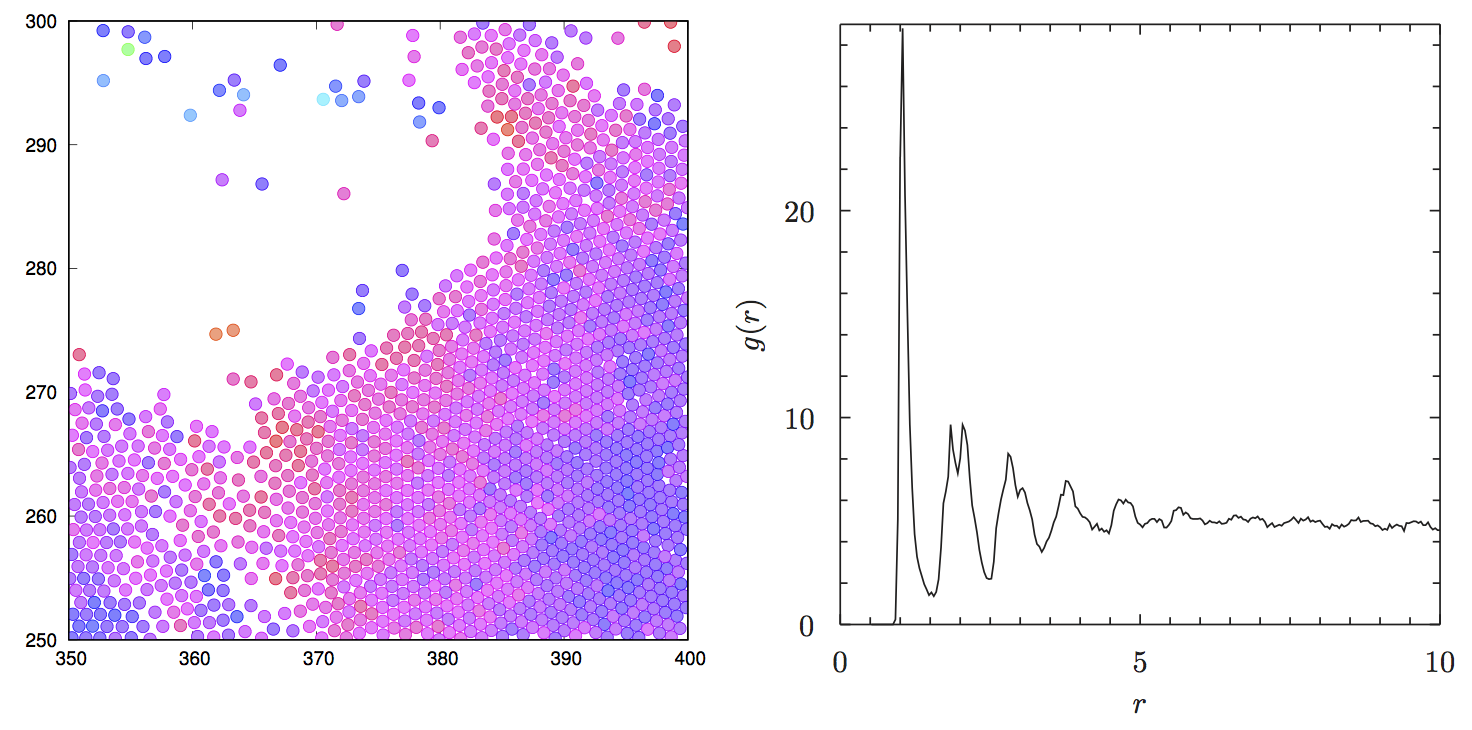}
\caption{Structure of the macro-drop. Left: Detailed view of the configuration shown in Fig. \ref{fig:snaps} (a), i.e. for $N=16000$, Pe$=10$, $g\rho=3$ and $\Omega=0.2$. 
Right: Corresponding pair-correlation function. }
\label{fig:strucPS}
\end{center}
\end{figure}

In order to further characterize the structure of the system in the macro-drop regime, we compute the pair-correlation function defined as
\begin{equation}
N\rho g(r)=\langle \sum_{i\neq j}\delta(|\boldsymbol{r}-\boldsymbol{r}_j+\boldsymbol{r}_i |)\rangle \, .
\end{equation}
This function, together with a detailed view of a configuration snapshot corresponding to the same set of parameters, is shown in Fig.~\ref{fig:strucPS}.
As shown in the snapshot and confirmed by the pair-correlation function, particles in the drop are closed packed, showing a highly ordered solid-like structure. 
Such a compact structure in the absence of attractive interactions is reminiscent to the dense phase generated by MIPS \cite{Redner2013, Levis2017}. Here, however, 
it occurs well below the expected critical point for MIPS. The underlying mechanism roots in the presence of alignment interactions 
which generate a coherently  moving structure. Excluded volume effects introduce an extra packing constraint on top of a compact structure which does not need volume interactions to emerge.    

\section{Micro-flock regime}

In the absence of excluded volume interactions, the characteristic size of the micro-flock pattern in the high frequency regime, is proportional to the radius of the single particle trajectory $l^\ast \propto$Pe$/\Omega$. 
This behavior has been predicted by the analysis of a coarse-grained hydrodynamic description and has been verified using particle based numerical simulations \cite{CAP}. 
Here, we explore the impact of excluded volume interactions among particles on these characteristic scaling laws following a careful analysis of the numerical simulations data.  
As a key result, we find that while excluded volume interactions change the size and in particular also the shape of these patterns, remarkably, the characteristic scaling law $l^\ast \propto$Pe$/\Omega$
still holds true.


\begin{figure}[h]
\begin{center}
\includegraphics[scale=0.17,angle=0]{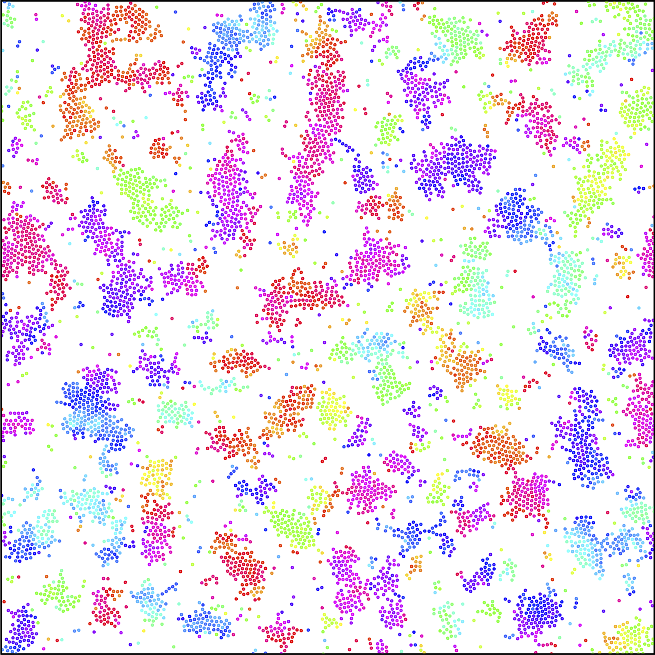}
\includegraphics[scale=0.17,angle=0]{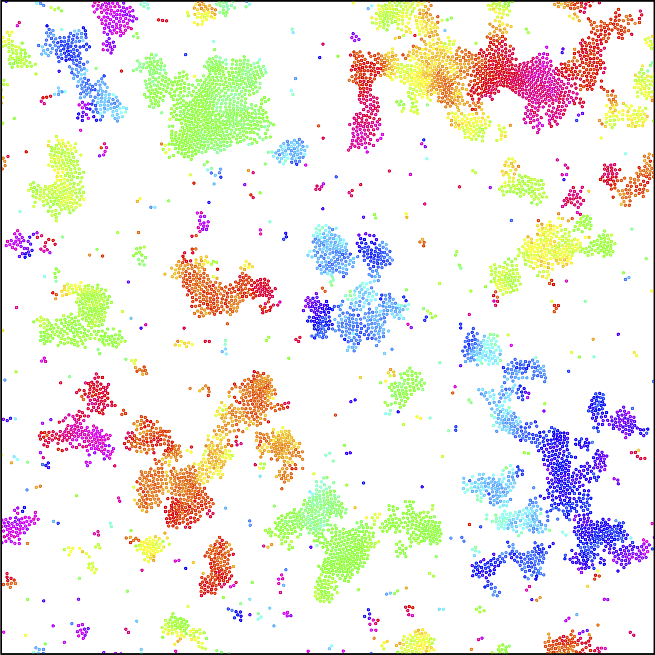}
\includegraphics[scale=0.17,angle=0]{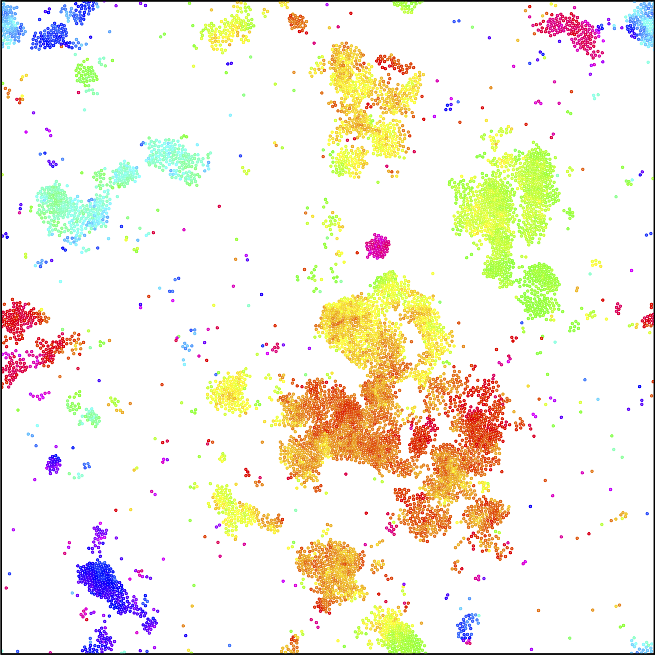}
\caption{Steady-state configurations  of $N=8000$ chiral active disks for  $g\rho=20$ and (from left to right) Pe$=1$, Pe$=5$ and Pe$=20$. The color map corresponds to the swimmers' 
orientation $\{\theta_i\}$. The size of the micro-flocks grows as the self-propulsion increases. }\label{fig:snapMF}
\end{center}
\end{figure} 

\begin{figure}[h]
\begin{center}
\includegraphics[scale=0.5,angle=0]{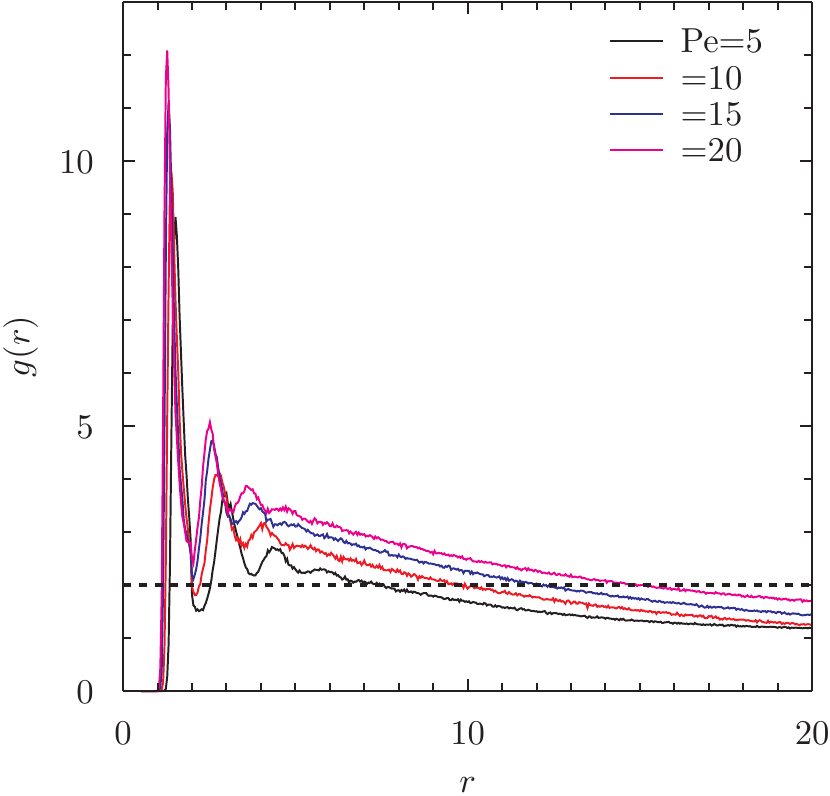}
\includegraphics[scale=0.5,angle=0]{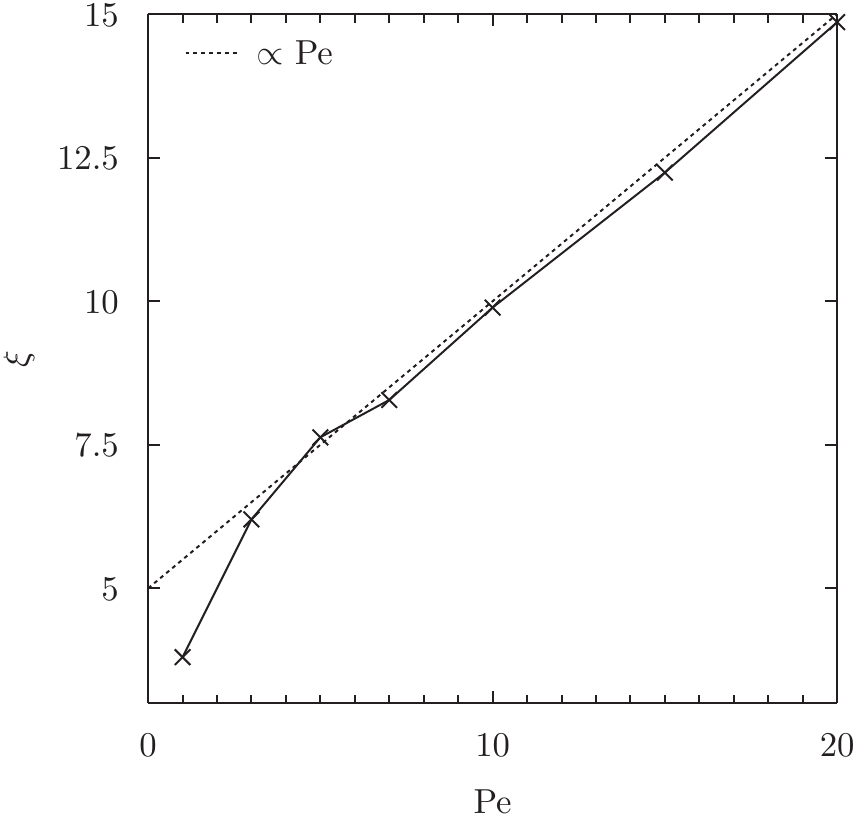}
\caption{Pair-correlation function of a system of  $N=8000$ chiral active disks at fixed  $\Omega=3$ and $g\rho=20$ for several P\'eclet numbers shown in the key (left). 
Characteristic length-scale extracted from the pair-correlation function data shown in the left panel as a function of the 
P\'eclet number (right).} \label{fig:xiPe}
\end{center}
\end{figure}

\begin{figure}[h]
\begin{center}
\includegraphics[scale=0.7,angle=0]{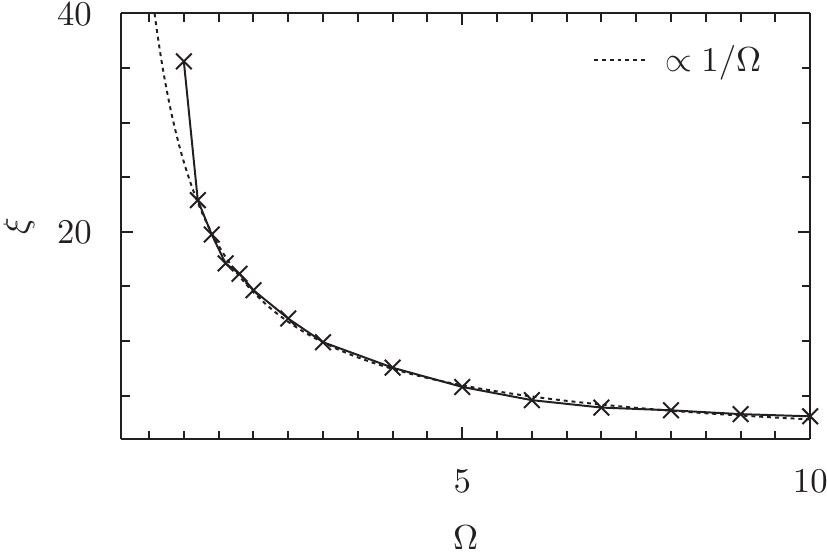}
\caption{Characteristic micro-flock length-scale as a function of the rotational frequency  at fixed  Pe$=10$ and $g\rho=20$ for $N=8000$.} \label{fig:xiO}
\end{center}
\end{figure}

In Fig.~\ref{fig:snapMF} we show snapshots of three systems of chiral active disks at different P\'eclet numbers (Pe$=1,5,20$). From inspection of the snapshots it is evident that the size 
of the micro-flock patterns increases with Pe (as for point-like circle swimmers \cite{CAP}). 
In order to define a characteristic length scale associated to these density heterogeneities, we use a criterion based on the analysis of the pair-correlation function rather that the one provided by 
the radius of gyration of the largest cluster used in the previous section. We define the length scale $\xi$ as the value above which the pair-correlation function is below some threshold, here, $g(r)<2$, $\forall\,r>\xi$. 
This quantity, which estimates  the range over which density correlations decay,
is illustrated in Fig. \ref{fig:xiPe} (a). As shown in the figure,  density correlations establish over larger length scales as  Pe increases, in agreement with the snapshots. 
 The results shown in Fig. \ref{fig:xiPe} (b) demonstrate the scaling law $\xi\propto$Pe  and  Fig. \ref{fig:xiO} confirms  $\xi\propto\Omega^{-1}$. 

\begin{figure}[h]
\begin{center}
\includegraphics[scale=0.45,angle=-0]{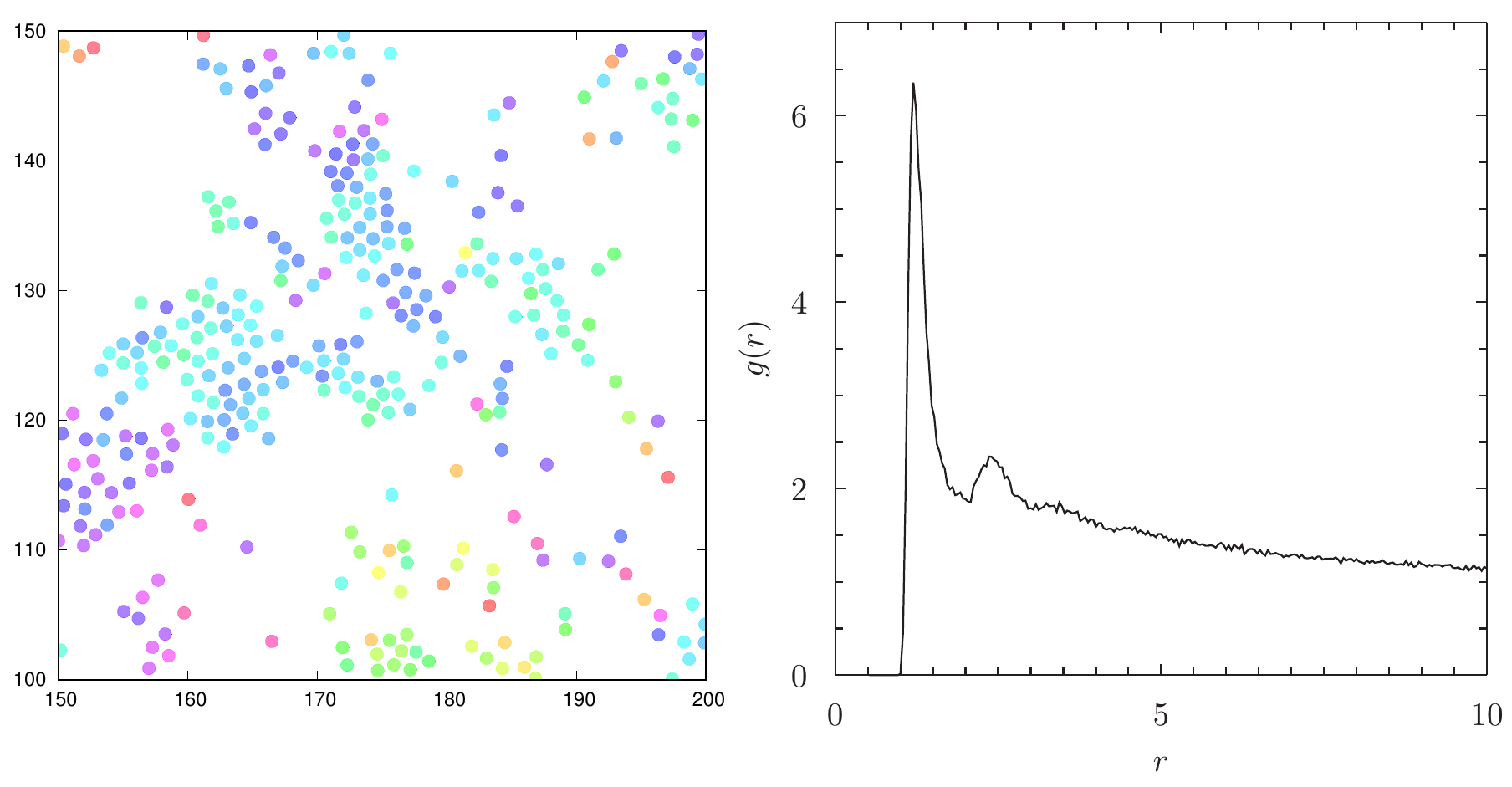}
\caption{Structure of the micro-flock. Left: Detailed view of the configuration shown in Fig. \ref{fig:snaps} (b) ($N=16000$, $g\rho=3$, $\Omega=3$). Right: Corresponding pair-correlation function for the  same system.}
\label{fig:strucMF}
\end{center}
\end{figure}

At a given coupling parameter above the flocking threshold, the micro-flocks show a less regular structure than their macro-drop counterparts. Indeed, fast rotations 
frustrate the tendency of the particles to accommodate their swim direction, resulting in smaller structures with a smaller degree of coherence. As shown in Fig.~\ref{fig:strucMF}, 
micro-flocks display a liquid-like structure rather than a solid-like one. 

\section{Conclusion}
The present work has shown that the typical collective behavior of chiral active matter, in the form of rotating macro-drop and micro-flock patterns, 
essentially survives the presence of excluded volume interactions among particles. 
In particular, despite strongly enhanced fluctuations of shape and size of individual structures, 
the 
characteristic linear scaling of the average micro-flock length-scale with the single-circle-swimmer-radius persists in presence of excluded volume interactions.
These findings might help to realize rotating macro-drop and micro-flock patterns experimentally using active colloids or biological circle swimmers and to test the 
predicted scaling law. 

\section{Acknowledgments}
DL and BL acknowledge received funding from the European Union's Horizon 2020 research and innovation programme under the Marie Sklodowska-Curie grant agreement No 657517 and No 654908. 
DL thanks Albert D\'iaz-Guilera, Aitor Mart\'in-Gomez, Elena Ses\'e and Ignacio Pagonabarraga for their collaborations in related topics.

\bibliographystyle{unsrt}
\bibliography{CAPab}

\begin{thebibliography}{10}

\bibitem{Berg1990}
Howard~C Berg and Linda Turner.
\newblock Chemotaxis of bacteria in glass capillary arrays. escherichia coli,
  motility, microchannel plate, and light scattering.
\newblock {\em Biophys. J.}, 58(4):919, 1990.

\bibitem{diLuzio2005}
W~R DiLuzio, L~Turner, M~Mayer, P~Garstecki, Douglas~B Weibel, H~C Berg, and
  G~M Whitesides.
\newblock Escherichia coli swim on the right-hand side.
\newblock {\em Nature}, 435(7046):1271, 2005.

\bibitem{Lauga2006}
E~Lauga, W~R DiLuzio, G~M Whitesides, and H~A Stone.
\newblock Swimming in circles: motion of bacteria near solid boundaries.
\newblock {\em Biophys. J.}, 90(2):400, 2006.

\bibitem{DiLeonardo2011}
R~Di~Leonardo, D~Dell’Arciprete, L~Angelani, and V~Iebba.
\newblock Swimming with an image.
\newblock {\em Phys. Rev. Lett.}, 106(3):038101, 2011.

\bibitem{Shenoy2007}
V~B Shenoy, D~T Tambe, A~Prasad, and J~A Theriot.
\newblock A kinematic description of the trajectories of listeria monocytogenes
  propelled by actin comet tails.
\newblock {\em Proc. Natl. Acad. Sci.}, 104(20):8229, 2007.

\bibitem{Friederich2012}
B~M Friedrich and F~J{\"u}licher.
\newblock Chemotaxis of sperm cells.
\newblock {\em Proc. Natl. Acad. Sci. U S A}, 104(33):13256, 2007.

\bibitem{Riedel2005}
I~H Riedel, K~Kruse, and J~Howard.
\newblock A self-organized vortex array of hydrodynamically entrained sperm
  cells.
\newblock {\em Science}, 309(5732):300, 2005.

\bibitem{Loewen2016}
H~L{\"o}wen.
\newblock Chirality in microswimmer motion: From circle swimmers to active
  turbulence.
\newblock {\em Eur. Phys. J. Spec. Top.}, 225(11):2319, 2016.

\bibitem{Kummel2013}
F~K{\"u}mmel, B~ten Hagen, R~Wittkowski, I~Buttinoni, R~Eichhorn, G~Volpe,
  H~L{\"o}wen, and C~Bechinger.
\newblock Circular motion of asymmetric self-propelling particles.
\newblock {\em Phys. Rev. Lett.}, 110(19):198302, 2013.

\bibitem{Hagen2014}
B~Ten~Hagen, F~K{\"u}mmel, R~Wittkowski, D~Takagi, H~L{\"o}wen, and
  C~Bechinger.
\newblock Gravitaxis of asymmetric self-propelled colloidal particles.
\newblock {\em Nature Commun.}, 5:4829, 2014.

\bibitem{Denk2016}
J~Denk, L~Huber, E~Reithmann, and E~Frey.
\newblock Active curved polymers form vortex patterns on membranes.
\newblock {\em Phys. Rev. Lett.}, 116(17):178301, 2016.

\bibitem{Kaiser2013}
A~Kaiser and H~L{\"o}wen.
\newblock Vortex arrays as emergent collective phenomena for circle swimmers.
\newblock {\em Phys. Rev. E}, 87(3):032712, 2013.

\bibitem{Liebchen2016}
B~Liebchen, M~E Cates, and D~Marenduzzo.
\newblock Pattern formation in chemically interacting active rotors with
  self-propulsion.
\newblock {\em Soft Matter}, 12:7259, 2016.

\bibitem{CAP}
B~Liebchen and D~Levis.
\newblock Collective behavior of chiral active matter: Pattern formation and
  enhanced flocking.
\newblock {\em Phys. Rev. Lett.}, 119(5):058002, 2017.

\bibitem{Vicsek1995}
T~Vicsek, A~Czir{\'o}k, E~Ben-Jacob, I~Cohen, and O~Shochet.
\newblock Novel type of phase transition in a system of self-driven particles.
\newblock {\em Phys. Rev. Lett.}, 75(6):1226, 1995.

\bibitem{Buttinoni2013}
I~Buttinoni, J~Bialk{\'e}, F~K{\"u}mmel, H~L{\"o}wen, C~Bechinger, and T~Speck.
\newblock Dynamical clustering and phase separation in suspensions of
  self-propelled colloidal particles.
\newblock {\em Phys. Rev. Lett.}, 110(23):238301, 2013.

\bibitem{Redner2013}
G~S Redner, M~F Hagan, and A~Baskaran.
\newblock Structure and dynamics of a phase-separating active colloidal fluid.
\newblock {\em Phys. Rev. Lett.}, 110(5):055701, 2013.

\bibitem{Stenhammar2014}
J~Stenhammar, D~Marenduzzo, R~J Allen, and M~E Cates.
\newblock Phase behaviour of active brownian particles: the role of
  dimensionality.
\newblock {\em Soft Matter}, 10(10):1489, 2014.

\bibitem{Cates2015}
M~E Cates and J~Tailleur.
\newblock Motility-induced phase separation.
\newblock {\em Annu. Rev. Cond. Matt. Phys.}, 6(1):219, 2015.

\bibitem{Levis2017}
D~Levis, J~Codina, and I~Pagonabarraga.
\newblock Active brownian equation of state: metastability and phase
  coexistence.
\newblock {\em arXiv preprint arXiv:1703.02412}, 2017.

\bibitem{VicsekRev}
T~Vicsek and A~Zafeiris.
\newblock Collective motion.
\newblock {\em Phys. Rep.}, 517(3):71, 2012.

\bibitem{Toner1995}
J~Toner and Y~Tu.
\newblock Long-range order in a two-dimensional dynamical xy model: how birds
  fly together.
\newblock {\em Phys. Rev. Lett.}, 75(23):4326, 1995.

\bibitem{Toner2005}
J~Toner, Y~Tu, and S~Ramaswamy.
\newblock Hydrodynamics and phases of flocks.
\newblock {\em Ann. Phys.}, 318(1):170, 2005.

\bibitem{Chate2008}
H~Chat{\'e}, F~Ginelli, G~Gr{\'e}goire, and F~Raynaud.
\newblock Collective motion of self-propelled particles interacting without
  cohesion.
\newblock {\em Phys. Rev. E}, 77(4):046113, 2008.

\bibitem{Farrell2012}
F~D~C Farrell, M~C Marchetti, D~Marenduzzo, and J~Tailleur.
\newblock Pattern formation in self-propelled particles with density-dependent
  motility.
\newblock {\em Phys. Rev. Lett.}, 108(24):248101, 2012.

\bibitem{SM}
{\em See Supplementary Material at doi:...}

\bibitem{Loose2013}
M~Loose and T~J Mitchison.
\newblock The bacterial cell division proteins ftsa and ftsz self-organize into
  dynamic cytoskeletal patterns.
\newblock {\em Nat. Cell Biol.}, 16(1):38, 2014.

\end{thebibliography}

\end{document}